\documentclass[12pt,aps,amsmath,latexsym,amsfonts]{JHEP3}
\usepackage{amsmath, amssymb, mathrsfs,graphics, graphicx}
\usepackage{epsfig}
\usepackage{amscd}
\usepackage[numbers, square, sort&compress]{natbib}

\title{Spinflation with backreaction}
\author{Dariush Kaviani\thanks{Email: dariush.kaviani@durham.ac.uk} \\
	Institute of Particle Physics Phenomenology, Department of Physics, \\ 
	Durham University, South Road,  Durham, DH1 3LE, UK}

\abstract{We study brane inflation in flux compactifications at the nonlinear
level, solving the D3-brane DBI-equations of motion including backreaction 
from the D3-D7 brane potential and from perturbations of the
warp factor. We first numerically compute the exact functional form of
the K\"ahler modulus valid on the entire supergravity background and
obtain a two-field potential along radial and harmonic directions. We
find that a valid perturbative expansion on the entire supergravity
background with the K\"ahler modulus integrated out adiabatically in DBI
inflation requires hierarchies of scales that determine
compactification parameters different from those typical in slow-roll
models. Our numerical results then show that the DBI inflationary
solutions are quite robust against these nonlinear corrections.}

\preprint{IPPP/12/95\\DCPT/12/190}

\begin{document}

\section{Introduction}

The isotropy and homogeneity of the universe suggests that in very
early times after the big bang the universe must have gone through
a period of rapid exponential expansion called inflation. The quest
for deriving inflation from a fundamental theory has led to the 
construction of inflationary models within the framework of string
theory. For embedding inflation into string theory, one has to 
specify a string compactification whose low-energy effective theory
contains a suitable inflaton field and potential. The many moduli 
fields that arise from Calabi-Yau compactifications can provide a 
large number of candidate scalar fields in the four-dimensional 
theory and any of these scalar fields may play the role of the 
inflaton field. However, moduli fields are generically either 
massless, or have a potential with runaway behaviour, which makes
their interpretation as inflatons rather difficult. In addition 
to this, the computation of the effective potential in terms of all 
these scalar fields and degrees of freedom is a highly nontrivial 
task. Nevertheless, by considering the systematics of flux 
compactifications \cite{Giddings:2001yu} and nonperturbative 
effects \cite{Kachru:2003aw} in string theory, it is possible to 
reduce the degrees of freedom and stabilize all the moduli fields 
as required for a suitable model of inflation.

One particular way of constructing string inflationary models has 
been within the `KKLT' framework, \cite{Kachru:2003aw}, according
to which the older idea of brane inflation \citep{Dvali:1998pa,Alexander:2001ks,Burgess:2001fx,Shiu:2001sy}      
is realised via the motion of a brane in the compact extra 
dimensions \cite{Kachru:2003sx}. It was shown in \cite{Kachru:2003sx} 
that the potential of the D3-brane in a Calabi-Yau flux compactification
of type IIB theory \cite{Giddings:2001yu} with the K\"ahler moduli 
stabilized according to \cite{Kachru:2003aw} receives corrections 
that make the inflaton potential too steep for slow-roll inflation. 
It was also conjectured in \cite{Kachru:2003sx} that further 
compactification corrections to the inflaton potential may lead to 
a finely tuned cancellation of correction terms, so that the inflaton 
potential can be made flat. Major progress in this direction was made
when these corrections were computed from ultraviolet deformations of
the warped throat geometry and applied to explicit brane inflation models \citep{Firouzjahi:2003zy,Burgess:2006cb,Chen:2008au,Chen:2008ada,Chen:2008ai,Baumann:2006th,Baumann:2007np,Baumann:2007ah,Baumann:2008kq,Baumann:2009qx,Baumann:2010ll,Kecskemeti:2006cg,DeWolfe:2007hd,Pajer:2008uy,Chen:2010qz} (see also
the reviews \citep{HenryTye:2006uv,Cline:2006hu,Kallosh:2007ig,Burgess:2007pz,Baumann:2009ds,Baumann:2009ni,McAllister:2007bg}). 
In these models the flattening of the inflaton potential is analysed
under fine tuning and vast scanning of the compactification parameters. 
In particular, in the `delicate universe' models studied in \citep{Chen:2008ai,Baumann:2007np,Baumann:2007ah}
it was shown that the inflaton potential can be made flat only in 
small region around an inflection point; elsewhere the potential 
remains steep. However, whether or not the potential can be made flat
to realise slow-roll conditions, inflation can occur by the DBI effect
 with high speed and steep potentials \cite{Silverstein:2003hf,Alishahiha:2004eh}.
But the usual approach taken by the DBI brane inflation models in the 
literature is to consider brane motion in an imaginary self-dual (ISD) 
compactification in which the effects of moduli stabilization and 
backreaction are not included. 

Any successful brane inflation model has to include the effects of 
compactification and moduli stabilization, which induce important 
corrections to the inflaton action. One particular correction from
moduli stabilization is the departure of the ISD solution which 
induces harmonic dependence in the action of the probe D3-brane, \citep{Baumann:2006th,Baumann:2007np,Baumann:2007ah,Baumann:2008kq,Baumann:2009qx,Baumann:2010ll}, 
which has been studied to some extend in \citep{Kecskemeti:2006cg,DeWolfe:2007hd,Pajer:2008uy}
and analysed in more detail in \cite{Gregory:2011cd}. 
In previous work \cite{Gregory:2011cd}, we solved the DBI brane 
equations of motion in the warped deformed conifold \cite{Klebanov:2000hb}
with harmonic dependent corrections from linearized perturbations around
the ISD solution. We showed that just as angular motion increases the number
of e-foldings (spinflation) \cite{Easson:2007dh,Easson:2009kk,Easson:2009wc,Cai:2010wt}, 
having additional angular dependence from linearized corrections also 
increases the number of e-foldings. However, this line of analysis 
considered multifield effects in D-brane inflation from only linearized
perturbations around the ISD supergravity solution. 
In this paper, we extend our previous analysis \cite{Gregory:2011cd} by
including further harmonic dependent corrections from non-linear 
perturbations around the ISD solution, which also contribute to the 
inflaton action. In the noncompact limit, non-linear perturbations 
are dominated by imaginary anti-self-dual (IASD) fluxes sourced by moduli
stabilizing wrapped D7-branes and the flux induced potential for the 
probe D3-brane in ten-dimensional supergravity equals the 
nonperturbatively-generated D3-D7 potential in four-dimensional 
supergravity \cite{Baumann:2010ll}.

One particular motivation of extending our previous analysis 
\cite{Gregory:2011cd} by including such nonperturbative corrections
is the possibility of increasing the amount of inflation and 
decreasing the level of non-Gaussianity due to backreaction effects \cite{Baumann:2009qx}.
The backreaction on the mobile D3-brane is sourced by itself \cite{Baumann:2006th}.
The D3-brane in a flux compactification containing holomorphically 
embedded D7-branes corrects the warp factor which in turn corrects
the warped volume of four-cycles wrapped by D7-branes. This then 
corrects the D3-D7 potential causing the backreaction on the D3-brane.
The perturbations of the warp factor are given by the Green's function
and correct the $\gamma_{\text{DBI}}$-factor which controls the level of 
non-Gaussianity. The Green's function solves the noncompact supergravity 
equation of motion describing the perturbations around the ISD solution 
sourced by IASD fluxes and may be expanded in an infinite set
of eigenstates of the Laplacian containing harmonic dependent 
hypergeometric functions \cite{Klebanov:2007us,Pufu:2010ie}.

In this paper we solve the D3-brane equations of motion from the
DBI action in the warped deformed conifold \cite{Klebanov:2000hb} 
with harmonic dependence from the leading correction terms of the
warp factor and the D3-D7 potential. To solve the brane equations
of motion, we note that the D3-brane potential including nonperturbative
corrections depends on the functional form of the K\"ahler modulus 
and that of the D7-brane embedding. For the Kuperstein embedding
of D7-branes \cite{Kuperstein:2004hy} and a simple choice of an 
harmonic dependent trajectory on the deformed conifold, we integrate
out the K\"ahler modulus and reduce the multifield D3-brane potential 
to a simple two-field potential depending on one radial and one 
harmonic direction of the conical geometry. In our numerical 
integrations we integrate out the K\"ahler modulus by computing
its exact functional form valid on both the infrared (IR) and 
ultraviolet (UV) regions of the supergravity background. We find
that computing the K\"ahler modulus within the adiabatic approximation
in DBI inflation requires certain hierarchies of scales which determine
the set of compactification parameters different from those in slow-roll
models. Integrating the brane equations of motion for the consistent 
choice of parameters with the numerically computed K\"ahler modulus 
shows that the DBI inflationary solutions are quite robust against
non-linear harmonic dependent corrections from perturbations of the
warp factor and the D3-D7 brane potential.

The paper is organized as follows. In section 2, we outline our 
supergravity background and specialize to the warped deformed conifold.
In section 3, we consider the general set up for brane inflation in the
warped deformed conifold deriving the explicit form of our D3-brane 
potential and its relating DBI brane equations of motion. In section 4,
we present our numerical results. We first determine the relevant 
parameter space and compute the K\"ahler modulus. We then solve the
DBI brane equations of motion. In section 5, we conclude with a brief
discussion.

\section{Type IIB supergravity}

The setting that we choose for our brane inflationary analysis 
is the Calabi-Yau flux compactification of type IIB theory 
containing a warped throat  region generated by fluxes \cite{Giddings:2001yu}.
We therefore consider backgrounds in low-energy IIB supergravity,
which in the Einstein frame can be represented by the action

\begin{align}
\label{2BEA}
S_{\,\text{IIB}} & =\frac{1}{2\kappa_{10}^2}\Bigg\{\int d^{10}x 
\sqrt{|g|}\left[\mathcal{R}_{10}-\frac{|\,\partial\tau|^2}
{2(\text{Im}\tau)^2}
-\frac{|\,G_{3}|^2}{12\,\text{Im}\tau}-\frac{|\,\tilde{F_{5}}|^2}
{4\cdot5\,!}\right] \notag \\ 
&+\frac{1}{4i}\int \frac{C_{4}\wedge G_{3} \wedge G_{3}^*}
{\text{Im}(\tau)}+S_{\,\text{loc}}\Bigg\}.
\end{align}
Here $S_{\,\text{loc}}$ stands for localized contributions from 
D-brane and orientifold planes; $\tau=C_{0}+ie^{-i\phi}$ is the
axion-dilaton field; $G_{3}=F_{3}-\tau H_{3}$ is the combination
of the R--R and NS--NS three-form fluxes $F_{3}=dC_{2}$ and 
$H_{3}=dB_{2}$ (with $C_2$ and $B_2$ being the R--R and NS--NN 
two-forms, respectively); $C_4$ is the R--R four-form; 
$\tilde{F}_{5}=dC_{4}-\frac{1}{2}C_{2}\wedge H_{3}+\frac{1}{2}B_{2}\wedge F_{3}$ 
is the self-dual five-form, $\tilde{F}_{5}=\star_{10}\tilde{F}_{5}$, 
with $\star_{10}$ denoting the ten-dimensional Hodge-star operator; 
$\mathcal{R}_{10}$ is the ten-dimensional Ricci-scalar; 
$\kappa_{10}^2=\frac{1}{2}(2\pi)^7 {{\alpha}^{\prime}}^4 g_{s}^2$
is the ten-dimensional gravitational coupling.

In a flux compactification we may take the line element of the form

\begin{equation}
\label{10dmetric}
ds_{10}^2=h^{-1/2}(y)ds_4^2-h^{1/2}(y)ds_6^2,
\end{equation} 
where $h$ is the warp factor depending only on the internal coordinates
$y:=y^m$, $ds_6^2$ is the metric on the internal manifold and $ds_4^2$
is the four-dimensional flat metric, which will be the FRW metric for 
cosmological analysis.

Following \cite{Giddings:2001yu}, we take the self-dual five-form to
be given by

\begin{equation}
\label{5F2}
\tilde{F}_{5}=(1+\star_{10})\Big[d\alpha(y)\wedge dx^{0}\wedge 
dx^{1}\wedge dx^{2}\wedge dx^{3}\Big],
\end{equation}
in the Poincare invariant case, where $\alpha(y)$ is a function of
the internal coordinates. The Einstein equations and five-form Bianchi
identity then imply

\begin{equation}
\label{GKPconst1}
\Delta_{(0)}\Phi_{\pm}=\frac{e^{\Phi}}{24 h^2}|\,G_{\pm}|^2
+h|\nabla\Phi_{\pm}|+\mathcal{R}_{4}+\text{local}
\end{equation}
where $\Delta_{(0)}$ is the Laplacian with respect to the 
six-dimensional unperturbed Calabi-Yau metric $g_{mn}^{(0)}$, and

\begin{equation}
\label{ISDflux}
G_{\pm}\equiv(i\pm\star_{6})G_{3},\;\;\;\;\;\ 
\Phi_{\pm}\equiv\frac{1}{h}\pm\alpha.
\end{equation}
The equation of motion for the three-form flux is

\begin{equation}
\label{GKPconst2}
d\Lambda+\frac{i}{2}\frac{d\tau}{\text{Im}(\tau)}
\wedge(\Lambda+\bar{\Lambda})=0,
\end{equation}
where by definition

\begin{equation}
\label{IASDflux}
\Lambda\equiv \Phi_{+}G_{-}+\Phi_{-}G_{+}.
\end{equation}
For $G_{-} = 0$, i.e. $\star_{6}G_{3} = iG_{3}$ , the flux $G_{3}$
is imaginary self-dual (ISD), the background metric is Calabi-Yau
with the five-form flux given by $\alpha= h^{-1}$ \cite{Giddings:2001yu}. 
The background that satisfies these conditions is called ISD.  
The specific example of such a background that we are interested in is 
the warped deformed conifold \cite{Klebanov:2000hb}. The deformed conifold
is a noncompact and nonsingular Calabi-Yau three-fold in $\mathbb{C}^4$ 
defined by the following constraint equation:
\begin{equation}
\label{dconifold}
\sum_{A=1}^{4}({z_{A}})^2=\epsilon^2.
\end{equation}
where $\{z_A,A=1,2,3,4\}$ represent the local complex coordinates in 
$\mathbb{C}^4$, and $\epsilon$ is the deformation parameter which can
be made real by phase rotation. For vanishing $\epsilon$, Eq.\,(\ref{dconifold})
gives the singular conifold and describes a cone over a five-dimensional 
Einstein manifold $X_5$. For us the nonsingular limit is relevant in which
$X_5$ is the $\big[SU(2)\times SU(2)\big]/\,U(1)$ coset space $T^{1,1}$ of
topology $S^{2}\times S^{3}$ parametrised by a set of five Euler angles
$\Psi=\{\theta_{i},\varphi_{i},\psi\}$ with $0\leq \theta_{i} \leq \pi$,
$0\leq \varphi_{i} \leq 2\pi$, $0\leq \psi \leq 4\pi$ $(i=1,2)$,  
and the would-be singularity at the tip, $r=0$, is replaced by a blown-up
$S^3$ of $T^{1,1}$ amounting to the deformation measured by $\epsilon$. 
The base of the cone can be parametrised   by the coordinates $y_{i}$ in 
a standard way \cite{Candelas:1989js}

\begin{eqnarray} 
\label{1_g8}                                                                 
y_{1}&=& \frac{1}{\sqrt{2}}\bigg(\cos\frac{\theta_{1}}{2}
\cos\frac{\theta_{2}}{2}e^{\frac{i}{2}\left(\varphi_{1}+
\varphi_{2}+\psi\right)}-\sin\frac{\theta_{1}}{2}
\sin\frac{\theta_{2}}{2}e^{-\frac{i}{2}\left(\varphi_{1}
+\varphi_{2}-\psi\right)}\bigg),\\ y_{2}&=& \frac{i}{\sqrt{2}}
\bigg(\cos\frac{\theta_{1}}{2}\cos\frac{\theta_{2}}{2}e^{\frac{i}{2}
\left(\varphi_{1}+\varphi_{2}+\psi\right)}+\sin\frac{\theta_{1}}{2}
\sin\frac{\theta_{2}}{2}e^{-\frac{i}{2}\left(\varphi_{1}+\varphi_{2}
-\psi\right)}\bigg),\\  y_{3}&=& -\frac{1}{\sqrt{2}}
\bigg(\cos\frac{\theta_{1}}{2}\sin\frac{\theta_{2}}{2}
e^{\frac{i}{2}\left(\varphi_{1}-\varphi_{2}+\psi\right)}
+\sin\frac{\theta_{1}}{2}\cos\frac{\theta_{2}}{2}
e^{\frac{i}{2}\left(\varphi_{2}-\varphi_{1}+\psi\right)}\bigg),
\\ \label{y4} y_{4}&=& \frac{i}{\sqrt{2}}
\bigg(\cos\frac{\theta_{1}}{2}\sin\frac{\theta_{2}}{2}
e^{\frac{i}{2}\left(\varphi_{1}-\varphi_{2}+\psi\right)}
-\sin\frac{\theta_{1}}{2}\cos\frac{\theta_{2}}{2}e^{\frac{i}{2}
\left(\varphi_{2}-\varphi_{1}+\psi\right)}\bigg).
\end{eqnarray}
In terms of these, the coordinates of the deformed conifold in 
(\ref{dconifold}) can be written as

\begin{align}                                                                  
\label{1_g12}
z_1 & =\frac{\epsilon}{\sqrt{2}}(e^{\frac{\eta}{2}}y_{1}
+e^{-\frac{\eta}{2}}\overline{y}_1) ,\;\;\;\;\;\;\  
z_2=\frac{\epsilon}{\sqrt{2}}(e^{\frac{\eta}{2}}y_{2}
+e^{-\frac{\eta}{2}}\overline{y}_{2}), \notag \\
z_3 &= \frac{\epsilon}{\sqrt{2}}(e^{\frac{\eta}{2}}y_{3}
+e^{-\frac{\eta}{2}}\overline{y}_{3}) , \;\;\;\;\;\;\ 
z_4=\frac{\epsilon}{\sqrt{2}}(e^{\frac{\eta}{2}}y_{4}
+e^{-\frac{\eta}{2}}\overline{y}_{4}),
\end{align}
where $\eta$ is the `radial coordinate'. The six-dimensional
Calabi-Yau metric, $ds_6^2$, on the deformed conifold can be
obtained from the K\"ahler potential given as \cite{Candelas:1989js}

\begin{equation}
\label{KSTinf1}
k(\eta)=\frac{\epsilon^{4/3}}{2^{1/3}}\int_{0}^{\eta}
{d\eta^{\prime}[\sinh(2\eta^{\prime})-2\eta^{\prime}]^{1/3}}.
\end{equation}
Here we note that this K\"ahler potential is derived from the relation

\begin{equation}
\label{reta}
r^3=\sum_{A=1}^{4}|z_A|^2=\epsilon^2\cosh\eta,
\end{equation}
and the metric for the deformed conifold reads as \cite{Candelas:1989js}

\begin{eqnarray}
\label{dcmetric}
ds_6^2&=&\frac{1}{2}\epsilon^{4/3}K(\eta)\bigg[ \frac{1}{3K(\eta)^3}
\{d\eta^2+(\omega^{5})^2\}+\cosh^2\frac{\eta}{2}\bigg\{\frac{1}{2}
\sum_{i=1}^{4}(\omega^{i})^2+\sum_{i\neq j=1}^{4}\omega^{i}
\omega^{j}\bigg\}\notag\\ &&+ \sinh^2\frac{\eta}{2}
\bigg\{\frac{1}{2}\sum_{i=1}^{4}(\omega^{i})^2-
\sum_{i\neq j=1}^{4}\omega^{i}\omega^{j}\bigg\}\bigg],
\;\;\;\;\;\;\;\;\;\;\;\;\;\;\;\;\;\;\;\;\;\;\;\;\;\
\end{eqnarray}
where

\begin{align}                                                                  
\label{1_g12}
\omega^{1} & = -\sin\theta_{1} d\varphi_{1},\;\;\;\;\;\;\  
\omega^{2}= d\theta_{1}, \;\;\;\;\;\;\  \omega^{3}= 
\cos\psi \sin\theta_{2}d\varphi_{2}-\sin\psi d\theta_{2},\notag \\
\omega^{4} &= \sin\psi\sin\theta_{2}d\varphi_{2}+
\cos\psi d\theta_{2}, \;\;\;\;\;\;\ \omega^{5}= d\psi
+\cos\theta_{1}d\varphi_{1}+\cos\theta_{2}d\varphi_{2},
\end{align}
and

\begin{equation}
\label{KKS}
K(\eta) = \frac{(\sinh(2\eta)-2\eta)^{1/3}}{2^{1/3}\sinh\eta}.
\end{equation}
The proper radial coordinate which measures the actual distance
up the throat in the six-dimensional metric is given by

\begin{equation}
\label{rks}
r(\eta)=\frac{\epsilon^{2/3}}{\sqrt{6}}\int_{0}^{\eta}{\frac{dx}{K(x)}}.
\end{equation}
The warping in this background is induced by the presence of 
type IIB background fluxes including $M$ units of $F_3$ flux
through the cycle $A$ and $-K$ units of $H_3$ flux through 
the cycle $B$. Due to the presence of three-form fluxes the 
above background emerges as a solution to Einstein's equations
and the resulting supergravity background is described by a 
throat with its tip being located at a finite radial coordinate
$r_{\text{IR}}$ while at $r_{\text{UV}}$ the throat is glued 
into an unwarped bulk geometry. The warp factor is given in 
terms of the $\eta$ coordinate as \cite{Klebanov:2000hb}

\begin{eqnarray}
h_{\text{KS}}&=&2(g_{s}M\alpha^{\prime})^2\,\epsilon^{-8/3}
\,I(\eta),\\ \label{I} I(\eta)&\equiv&\int_{\eta}^{\infty}
{dx\frac{x\cosh x-1}{\sinh^2x}(\sinh x\cosh x-x)^{1/3}}.
\end{eqnarray}
At small and large radius, the warp factor takes the form:

\begin{equation}
\label{H2}
 h_{\text{KS}}=
    \begin{cases}
     2(g_{s}M\alpha^{\prime})^2\epsilon^{-8/3}
\bigg(0.5699-2\epsilon^{-4/3}\frac{r^2}{3}\bigg);
\,\,\, \eta\rightarrow \text{small}\nonumber\\
      \frac{27}{8}\frac{(g_{s}M\alpha^{\prime})^2}
{r^4}\bigg(\ln\frac{r^3}{\epsilon^2}+\ln\frac{4\sqrt{2}}{3\sqrt{3}}
-\frac{1}{4}\bigg);\;\;\;\;\;\;\;\; \eta\rightarrow \text{large}\nonumber
    \end{cases} 
\end{equation}

In an ISD compactification such as the warped deformed conifold,
only the complex-structure moduli are stabilized but not the 
K\"ahler moduli. Embedding the warped throat into the Calabi-Yau
sector with K\"ahler moduli stabilized breaks the no-scale 
structure and induces perturbations around the ISD solution. 
In the noncompact limit ($M_{\text{pl}}\rightarrow \infty$), 
these perturbations are sourced by non-linear effects including
only IASD fluxes. These fluxes satisfy the IASD conditions \cite{Baumann:2009qx}

\begin{equation}
\label{1_g20}
d\Lambda=0, \;\;\;\;\;\;\;\;\;\ \text{where}\;\;\;\;\;\;\;\;\;\;
\star_{6}\Lambda=-i\Lambda
\end{equation}
on general Calabi-Yau cones.
The perturbations of $\Phi_{-}$ around ISD conditions containing
IASD fluxes as the dominant source satisfy the second-order 
supergravity equation of motion \cite{Baumann:2009qx,Baumann:2010ll}

\begin{equation}
\label{1_g21}
\Delta_{(0)}\Phi_{-}=\frac{g_{s}}{96}|\Lambda|^2,
\end{equation}
where $\Delta_{(0)}$ is the Laplacian obtained from the unperturbed
Calabi-Yau metric $g_{mn}^{(0)}$, as before. The solution of 
Eq.\,(\ref{1_g21}) can always be written as the sum of first order
homogeneous and second order inhomogeneous solution in the form \cite{Baumann:2009qx,Baumann:2010ll}:

\begin{equation}
\label{1_g22}
\Phi_{-}(y)=\Phi_{-}^{(1)}(y)+\Phi_{-}^{(2)}(y),
\end{equation}
where the homogeneous solution satisfies the Laplace equation

\begin{equation}
\label{1_g24}
\Delta_{(0)}\Phi_{-}^{(1)}(y)=0.
\end{equation}
A particularly simple solution of the Laplace equation on the
deformed conifold takes the form \cite{Gregory:2011cd}

\begin{equation}
\label{KSefn}
\Phi_{-}^{(1)}(\eta, \theta)\propto (\cosh\eta\sinh\eta-\eta)^{1/3}\cos\theta.
\end{equation}
The inhomogeneous solution can always be written in the form:

\begin{equation}
\label{inhsol}
\Phi_{-}^{(2)}(y)=\frac{g_{s}}{96}\int{d^{\,6} y^{\prime}
\mathcal{G}(y;y^{\prime})|\Lambda|^2(y^{\prime})},
\end{equation}
where

\begin{equation}
\label{1_g23}
\Delta_{(0)} \mathcal{G}(y;y^{\prime})=\delta(y-y^{\prime}).
\end{equation}
Here and in what follows $y$ denotes a collective internal 
coordinate consisting of the radial and six angular directions,
where the later will always be denoted by $\Psi$.
On a general Calabi-Yau cone with K\"ahler form $J$ and 
holomorphic (3,0) form $\Omega$, we may turn on (1,2) flux,
$\Lambda_{1}$, and a non-primitive (2,1) flux, $\Lambda_{2}$ \cite{Baumann:2009qx,Baumann:2010ll}:

\begin{equation}
\label{1_g29}
\Lambda_{1}=\partial\partial f_{1}\bar{\Omega},
\;\;\;\;\;\;\;\;\;\;\;\ \Lambda_{2}=\partial f_{2}\wedge J,
\end{equation}
with $f_{1}$ and $f_{2}$ being holomorphic functions. 
The solution (\ref{inhsol}) takes the form \cite{Baumann:2009qx,Baumann:2010ll}:
\begin{equation}
\label{1_g30}
\Phi_{-}^{(2)}(y)=\frac{g_{s}}{32}\left[\mathcal{K}^{\Sigma\bar{\Xi}}
\partial_{\Sigma} f_{1}\overline{\partial_{\Xi}f_{1}}+2|f_{2}|^2\right],
\end{equation}
where $\mathcal{K}^{\Sigma\bar{\Omega}}$ is the K\"ahler metric. 
For a specific choice of the $f_i$'s, Eq.\,(\ref{1_g30}) determines
the explicit solution of Eq.\,(\ref{1_g21}) up to harmonic terms.

In the IR region of the throat where $\eta$ is small, the Green's 
function is that of the deformed conifold and takes the form \cite{Pufu:2010ie}

\begin{equation}
\label{1_g26}
\mathcal{G}(y;y^{\prime})=\mathcal{G}(\eta,y_{4};\bold{e_{0}})=
-\frac{3^{2/3}}{2^{8/3}\pi^3\epsilon^{8/3}}\sum_{j=0,\frac{1}{2},1,...}
\sum_{m=-j}^{j}\frac{\sqrt{2j+1}}{\eta}\mathcal{F}_{jm}(y_{4},\overline{y}_4),
\end{equation}
where $\bold{e_{0}}$ parametrizes the blown up $S^3$ at $\eta=0$
and $\mathcal{F}_{jm}$ stand for the hypergeometric functions given by

\begin{align}                                                                  
\label{1_g27}
\mathcal{F}_{0,0} &= 1,\;\;\;\;\;\;\  \mathcal{F}_{\frac{1}{2},
\frac{1}{2}}= 2y_{4},\;\;\;\;\;\;\  \mathcal{F}_{\frac{1}{2},
-\frac{1}{2}}= 2\overline{y}_4, \;\;\;\;\;\;\ \mathcal{F}_{1,1}=
2\sqrt{3}y_{4}^2 , \notag \\
\mathcal{F}_{1,0} &= -\sqrt{3}(1-4y_{4}\overline{y}_4),
\;\;\;\;\;\;\;\;\ \mathcal{F}_{1,-1}=2\sqrt{3}\overline{y}_4^2, 
\;\ \mathcal{F}_{\frac{3}{2},\frac{3}{2}}=4\sqrt{2}y_{4}^3, \notag \\
\mathcal{F}_{\frac{3}{2},\frac{1}{2}} &= -4\sqrt{2}y_{4}(1-3y_{4}
\overline{y}_4),\;\;\;\ \mathcal{F}_{\frac{3}{2},-\frac{1}{2}}=
-4\sqrt{2}\overline{y}_4(1-3y_{4}\overline{y}_4), \notag \\
\mathcal{F}_{\frac{3}{2},\frac{3}{2}} &=4\sqrt{2}\overline{y}_4^3.
\end{align}

In the UV region of the throat where $\eta$ is large, we may introduce
another radial coordinate $r$ through $r^3\sim\epsilon^2 e^{\eta}$ 
(see Eq.\,(\ref{reta})) and the Green's function is that of the singular
conifold given by \cite{Baumann:2006th}

\begin{equation}
\label{H2}
 \mathcal{G}(y;y^{\prime})=\sum_{L}\frac{Y_{L}^{*}(\Psi^{\prime})
 Y_{L}(\Psi)}{2\sqrt{\Lambda_{L}+4}}\times
    \begin{cases}
     \frac{1}{{r^{\prime}}^4} \bigg(\frac{r}{r^{\prime}}\bigg)^{c_{L}^{+}}
 & r\leq r^{\prime}  \\
      \frac{1}{r^4} \bigg(\frac{r^{\prime}}{r}\bigg)^{c_{L}^{+}}
 & r\ge r^{\prime}
    \end{cases} 
\end{equation}
with the harmonic eigenfunctions

\begin{equation}
\label{H3}
Y_{L}(\Psi)=Z_{j_1,m_1,R}(\theta_1)Z_{j_2,m_2,R}(\theta_2)
e^{im_1\varphi_1+im_2\varphi_2}e^{\frac{i}{2}R\psi},
\end{equation}

\begin{eqnarray}
\label{H4}
Z_{j_i,m_i,R}^{\text{I}}(\theta_i)&=&N_L^{\text{I}}(\sin\theta_i)^{m_i}
\bigg(\cot\frac{\theta_i}{2}\bigg)^{R/2}\times\notag \\ && {}_2F_{1}
\bigg(-j_i+m_i,1+j_i+m_i,1+m_i-\frac{R}{2};\sin^2\frac{\theta_i}{2}
\bigg),\\ \label{H5} Z_{j_i,m_i,R}^{\text{II}}(\theta_i)&=&
N_L^{\text{II}}(\sin\theta_i)^{R/2}\bigg(\cot\frac{\theta_i}{2}\bigg)^{m_i}\times
\notag \\ && {}_2F_{1}\bigg(-j_i+\frac{R}{2},1+j_i+\frac{R}{2},1-m_i+\frac{R}{2};
\sin^2\frac{\theta_i}{2}\bigg).
\end{eqnarray}
The normalization factors $N_L^{\text{I}/\text{II}}$ relation is given by

\begin{equation}
\label{norm}
V_{T^{1,1}}\int_0^1{dx [Z_{j_1,m_1,R}(x)]^2}\int_0^1{dy [Z_{j_2,m_2,R}(y)]^2}=1.
\end{equation}
In the above relations ${}_2F_{1}(a,b,c;d)$ stands for hypergeometric
functions; $L$ is a multi-index with the data $L\equiv (j_1,j_2),
(m_1,m_2),R$, where $j_1$ and $j_2$ are both integers or half-integers
with $m_1\in\{-j_1,\cdots,j_1\}$ and $m_2\in\{-j_2,\cdots,j_2\}$; 
$\Lambda_{L}=6(j_1(j_1+1)+l_2(j_2+1)-R^2/8)$ denotes the spectrum of
the full wave function and the eigenfunctions transform under 
$SU(2)_1\times SU(2)_2$ as the spin $(j_1,j_2)$ representation and 
under the $U(1)_R$ with charge $R$; $c^{+}_{L}\equiv -2\pm \sqrt{\Lambda_L+4}$.

\section{Multifield D-brane inflation}

We now embed a mobile D3-brane in the supergravity background
described in the previous section and analyse its the 
four-dimensional effective action. In the supergravity background
with metric ansatz (\ref{10dmetric}) the effective action takes
the form

\begin{eqnarray}
\label{DIF2}
I &=& \frac{M_{Pl}}{2}\int{d^4x\sqrt{-g}\mathcal{R}} 
-g_{s}^{-1}\int{d^4 x\sqrt{-g}\bigg[h^{-1}
(\gamma_{\text{DBI}}^{-1}-1) +V(\phi^m)\bigg]},
\\ \gamma_{\text{DBI}}^{-1} &=& \sqrt{1-hg_{mn}g^{\mu\nu}
\partial_{\mu}\phi^m\partial_{\nu}\phi^n}\nonumber.
\end{eqnarray}
Here $g_s$ is the string coupling, and $M_{\text{pl}}$ is the
Planck-mass. The first term in this action is the ordinary 
four-dimensional Einstein-Hilbert action, which arises from 
dimensional reduction of the closed string sector of the 
ten-dimensional action. The second part contains the action 
that controls the dynamics of the fields, parametrizing the 
position of the brane along the internal coordinates, $\phi^m$.
In a strongly warped region, $h\gg 1$, the kinetic energy's
pre-factor of $h^{-1}$ in Eq. (\ref{DIF2}) suppresses it 
relative to $V(\phi^m)$ even when the motion is relativistic.

To study brane inflation, we will take the position of the 
D3-brane to be homogeneous,\ $\phi^m = \phi^m(t)$, and we will
consider the four-dimensional metric to be the standard 
unperturbed FRW metric:
\begin{equation}
ds_4^2=g_{\mu\nu} dx^\mu dx^\nu = dt^2 - a^2(t) d{\bf x}^2,
\end{equation}
where $a(t)$ is the scale factor. The variation of the later
part of the action (\ref{DIF2}) with respect to the metric 
produces the energy-momentum tensor, which has the form of
a perfect fluid with its energy density and pressure given by:

\begin{eqnarray}
E&=&T_3(h^{-1}[\gamma_{\text{DBI}}-1]+V],\\ P&=&T_3(h^{-1}
[1-\gamma_{\text{DBI}}^{-1}]+V].
\end{eqnarray}
Here $T_3=((2\pi)^3 {\alpha^{\prime}}^2)^{-1}$ is the D3-brane
tension. The full equations of motion are:

\begin{eqnarray}
\label{nonhub1}
H^2 &=& \frac{E}{3M_\text{pl}^2},  \\ \label{nonhub1der}
{\dot H} &=& -\frac{(E+P)}{2\,M_\text{pl}^2}
\label{ray1},\\ \label{EqfieldsNew}
\frac{1}{a^3} \frac{d}{dt}\left[ a^3 \gamma_{\text{DBI}}
 \,{ g_{mn}^{(0)}}\, {\dot \phi}^n
\right]  &=&  -T_3\gamma_{\text{DBI}}
\,\left(\gamma_{\text{DBI}}^{-1}- 1\right)^2
 \, \frac{\partial_m h}{2h^2}
+\frac{\gamma_{\text{DBI}}}{2} 
\,\frac{\partial { g}_{ln}^{(0)}}{\partial \phi^m}
\,{\dot \phi}^l {\dot \phi}^n
- \partial_m V, \notag\\ \label{eqsfields}\,
\end{eqnarray}
where $g_{mn}^{(0)}$ is the unperturbed Calabi-Yau metric, 
as before. The Planck-mass, $M_{\text{pl}}$, appearing above
is bounded by the UV-scale as \cite{Gregory:2011cd}:

\begin{equation}
\label{MP}
M_{\text{pl}}^2>\frac{\epsilon^{4/3}g_sM^2 T_3}{6\pi}J(\eta_{\text{UV}}),
\end{equation}
where $J(\eta)=\int{I(\eta)\sinh^2\eta}$ with $I(\eta)$ given by (\ref{I}).

The D3-brane potential appearing in the above equations of motion is:

\begin{equation}
\label{D3pot1}
T_3V=\frac{1}{2}m^2\phi^2+T_3\Phi_{-}.
\end{equation}
Here the first term is a mass term generated for the canonical
inflaton $\phi=\sqrt{T_3} r(\eta)$, which by (\ref{rks}) takes
the form:

\begin{equation}
\label{rks}
\phi=\epsilon^{2/3}\sqrt{\frac{T_3}{6}}\int_{0}^{\eta}{\frac{dx}{K(x)}}.
\end{equation}
We note that the normalization is strictly dependent on the position
of the brane, which affects the volume modulus (see e.g.\
\citep{Baumann:2006th,Baumann:2007np,Baumann:2007ah,Baumann:2008kq,Baumann:2009qx,Baumann:2010ll}),
and also on the trajectory of the brane, even in the slow roll
approximation, due to the inflaton, $\phi$, being a sigma model field,
\citep{Achucarro:2010da,Langlois:2008mn,Huang:2007hh}. The last term in 
(\ref{D3pot1}) arises from perturbations around the noncompact (ISD) 
supergravity solution described by Eq.\,(\ref{1_g21}). The homogeneous
linearized solution of Eq.\,(\ref{1_g21}) is given by (\ref{KSefn}). 
The inhomogeneous non-linear solution of Eq.\,(\ref{1_g21}) given in
its implicit form as (\ref{1_g30}) equals the nonperturbatively
generated D3-brane F-term potential in four-dimensional supergravity
\cite{Baumann:2010ll}. The F-term potential takes the following form:

\begin{equation}
\label{super8}
V_F=e^{\kappa^2\mathcal{K}}\left[D_{\Sigma}W
\mathcal{K}^{\Sigma\overline{\Xi}}\overline{D_{\Xi}W}-3\kappa^2 W
\overline{W}\right],
\end{equation}
where $\{z^{\Sigma}\}\equiv \{\rho, z_{\alpha}; \alpha=1,2,3\}$ and
$D_{\Sigma}W=\partial_{\Sigma}W+\kappa^2 (\partial_{\Sigma}\mathcal{K})W$
with $\mathcal{K}$ and $W$ denoting the K\"ahler potential and 
superpotential, and $\kappa^2=M_{\text{pl}}^{-2}$ with $M_{\text{pl}}^2$ 
bounded by (\ref{MP}). From (\ref{MP}) one can see that large 
$\eta_{\text{UV}}$ implies large $M_{\text{pl}}$. In the noncompact
limit ($M_{\text{pl}}\rightarrow \infty$), the F-term potential 
(\ref{super8}) takes the form

\begin{equation}
V_{F}=\mathcal{K}^{\Sigma\overline{\Xi}}\partial_{\Sigma}W
\overline{\partial_{\Xi}W}.
\end{equation}
For any holomorphic function $W$ on the conical geometry, 
we may turn on the flux 

\begin{equation}
\label{3flux}
\Lambda_{\Sigma\bar{\Xi}\bar{\Gamma}}=\partial_{\Sigma}
\partial_{\Upsilon}W \mathcal{K}^{\Upsilon\bar{\Theta}}
\bar{\Omega}_{\bar{\Theta}\bar{\Xi}\bar{\Gamma}}=
\partial\partial W\cdot \bar{\Omega},
\end{equation}
where $\Omega$ is the holomorphic $(3,0)$-form. 
Solving Eq.\,(\ref{1_g21}) for Eq.\,(\ref{3flux}) gives \cite{Baumann:2010ll}

\begin{equation}
T_{3}\Phi_{-}=\mathcal{K}^{\Sigma\bar{\Xi}}\partial_{\Sigma}W
\overline{\partial_{\Xi}W}.
\end{equation}
The K\"ahler potential, $\mathcal{K}$, depends on the complex K\"ahler
modulus $\rho=\sigma+i\chi$, and on the D3-brane position, 
$\{z_{\alpha},\bar{z}_{\alpha}\}$, \cite{DeWolfe:2002nn}

\begin{equation}
\label{super9}
 \mathcal{K}(z^{\alpha},\bar{z}^{\alpha},\rho,\bar{\rho}) =  
-3 \kappa^{-2}\log[\rho+\bar{\rho}-\gamma k(z^{\alpha},
\bar{z}^{\alpha})] \equiv -3\kappa^{-2}\log U(z,\rho).
\end{equation}
Here $k(z^{\alpha}, \bar{z}^{\alpha})$ is the so-called 
`little' K\"ahler potential of the Calabi-Yau manifold, 
$\kappa^2=M_{\text{pl}}^{-2}$ as before, and $\gamma$ 
is a normalization factor given by

\begin{equation}
\gamma=\frac{\sigma_{0}T_{3}}{3M_{\text{pl}}^2}
\end{equation}
where $\sigma_{0}$ denotes the value of $\sigma$ when
the D3-brane is at its stabilized configuration \cite{Firouzjahi:2003zy,Baumann:2007np}. 
The K\"ahler metric and its inverse take the form \cite{Firouzjahi:2003zy}

\begin{eqnarray}
\label{super10}
\mathcal{K}_{\Xi\overline{\Sigma}} &=& \frac{3}{\kappa^{2}U^{2}}
\left(\begin{tabular}{c|c} $1$ & $-\gamma k_{\bar{\beta}}$\\ 
\hline $-\gamma k_{\alpha}$ & $U \gamma k_{\alpha \bar{\beta}}
+\gamma^{2} k_{\alpha}k_{\bar{\beta}}$\end{tabular}\right),
\\  \label{super11} \mathcal{K}^{\Delta\overline{\Gamma}} &=&
\frac{\kappa^{2}U}{3}\left(\begin{tabular}{c|c} $U
+\gamma k_{\gamma} k^{\gamma \bar{\delta}}k_{\bar{\gamma}}$ 
& $k_{\gamma}k^{\gamma \bar{\beta}}$ \\ \hline 
$k^{\alpha \bar{\delta}}k_{\bar{\delta}}$ & $\frac{1}{\gamma} 
k^{\alpha \bar{\beta}}$\end{tabular}\right),
\end{eqnarray}
where $k_{\alpha\bar{\beta}} \equiv \partial_{\alpha}
\partial_{\bar{\beta}}k$ is the Calabi-Yau metric, and 
$k_{\alpha} \equiv k_{,\alpha}$. The superpotential $W$,
also depends on the D3-brane positions, $\{z_{\alpha}\}$,
and is given by

\begin{equation}
\label{super12}
W(\rho)=W_0+A(z^{\alpha})e^{-a\rho}.
\end{equation}
The first part of the superpotential is the Gukov-Vafa 
perturbative superpotential \cite{Gukov:1999ya}. The 
second part in the superpotential comes from nonperturbative
effects sourced by moduli stabilizing D7-branes wrapping 
certain four-cycles in the compactification and $a=2\pi/n$ is
a constant with $n$ denoting the number of wrapped branes. The
inflaton dependence of the superpotential is induced by the 
interaction between the inflationary D3-brane and wrapped D7-brane.
The displacement of the D3-brane in the compactification slightly 
modifies the supergravity background, perturbing the warp factor, 
$h=h_{\text{KS}}+\delta h$, which corrects the warped volume of 
four-cycles from D7-branes. This then corrects the D3-D7 potential 
inducing the backreaction on the mobile D3-brane. 
The prefactor in the nonperturbative part of (\ref{super12}) 
computed from corrections to the warped background takes the form \cite{Baumann:2006th}:

\begin{equation}
\label{super7}
A(z^{\alpha})=A_{0}\bigg[\frac{f(z^{\alpha})}{f(0)}\bigg]^{1/n},
\end{equation}
where $f(z^{\alpha})$ denotes the holomorphic embedding
function of D7-branes depending on the D3-brane coordinates
and $A_0$ is a constant.

According to this and (\ref{super10})\,-\,(\ref{super11}), 
the scalar potential (\ref{super8}) takes the form \cite{Firouzjahi:2003zy,Baumann:2007np}

\begin{eqnarray}
\label{super13}
V_{F}(z^{\alpha},\bar{z}^{\alpha},\rho,\bar{\rho}) & = & 
\frac{\kappa^2}{3 U(z,\rho)^2}\bigg\{\Big[U(z,\rho) + 
\gamma k^{\gamma\bar{\delta}} k_{\gamma} k_{\bar{\delta}}\Big]
|W_{,\rho}|^2-3(\overline{W}W_{,\rho}+c.c.)\bigg\}\notag\\ && 
+\frac{\kappa^2}{3 U(z,\rho)^2}\bigg[\Big(k^{\alpha\bar{\delta}}
 k_{\bar{\delta}}\overline{W}_{,\bar{\rho}} W_{,\alpha} + c.c.
\Big)+\frac{1}{\gamma}k^{\alpha\bar{\beta}}W_{,\alpha}
\overline{W}_{,\bar{\beta}}\bigg].
\end{eqnarray}
The first part in Eq.\,(\ref{super13}) is the standard KKLT
F-term potential, and the second part arises exclusively from
corrections to the nonperturbative superpotential. By considering
appropriate formulas for the various terms in the potential 
(e.g. see \cite{Chen:2008ai}), it is straightforward to show that
the F-term potential (\ref{super13}) takes the functional form 
$V_{F}=V_{F}(z_1+\bar{z}_{1}, |z_1|^2, \eta,\sigma,\chi)$. To obtain
the explicit form of the potential, we need to specify an embedding
for D7-branes. For the Kuperstein embedding of D7-branes \cite{Kuperstein:2004hy},
 we have

\begin{equation}
f(z_1)=\mu-z_1=0.
\end{equation}
Thus the prefactor of the nonperturbative superpotential, 
Eq.\,(\ref{super7}), and its derivative with respect to the
independent coordinates take the form

\begin{equation}
\label{KTpot6}
A(z_1)=A_0\bigg(1-\frac{z_1}{\mu}\bigg)^{1/n},
\end{equation}

\begin{equation}
\label{KTpot7}
A_{i}(z_{1})=-\frac{A_{0}}{n\mu}\bigg(1-\frac{z_{1}}
{\mu}\bigg)^{1/n-1}\delta_{i1}.
\end{equation}
To obtain a simple trajectory depending only on one
 harmonic mode, say $\theta_1=\theta$, we fix the rest
 of angular directions of $T^{1,1}$ by imposing the 
following constrains

\begin{align}                                                                 
\label{nonlin2}
\frac{\varphi_{1}-\varphi_{2}\pm\psi}{2} & = \pm\frac{\pi}{2}
, \;\;\;\;\;\;\  \frac{\varphi_{1}+\varphi_{2}+\psi}{2} =\pi, \notag \\
\frac{\psi-\varphi_{1}-\varphi_{2}}{2} &=0, \;\;\;\;\;\;\;\;\;\;\ \theta_2=0.
\end{align}
Accordingly to this, the coordinates on the deformed conifold read as

\begin{align}                                                                  
\label{nonlin3}
z_{1} & = -\epsilon\cosh\bigg(\frac{\eta}{2}\bigg)
\cos\bigg(\frac{\theta}{2}\bigg), \;\;\;\;\;\;\  
z_{2} = -i\epsilon\sinh\bigg(\frac{\eta}{2}\bigg)
\cos\bigg(\frac{\theta}{2}\bigg), \notag \\
z_{3} & = -i\epsilon\sinh\bigg(\frac{\eta}{2}\bigg)
\sin\bigg(\frac{\theta}{2}\bigg), \;\;\;\;\;\;\ 
z_{4} = +\epsilon\cosh\bigg(\frac{\eta}{2}\bigg)
\sin\bigg(\frac{\theta}{2}\bigg).
\end{align}
The imaginary part of the K\"ahler modulus can be integrated out by

\begin{equation}
\label{intoutmod}
\frac{e^{ia\chi}}{A}\rightarrow\frac{1}{|A|}.
\end{equation}
Accordingly, the four-dimensional supergravity potential
is\footnote{It is straightforward to check that for $\theta=0$
our two-field potential (\ref{nonlin4}) coincides with the 
single field potential derived in \cite{Chen:2008ai}.}

\begin{eqnarray}
\label{nonlin4}
V_{F}&=&\frac{2\kappa^2a_{n}^2|A_{0}|^2e^{-2a_{n}\sigma}}{U^2}
|g(\eta,\theta)|^{2/n}\notag\\ &&\times\bigg\{\frac{U}{6} +
\frac{1}{a_{n}}\bigg(1-\frac{|W_{0}|}{|A_{0}|}
\frac{e^{a_{n}\sigma}}{g(\eta,\theta)^{1/n}}\bigg)
+F(\eta,\theta)\bigg\},
\end{eqnarray}
where

\begin{eqnarray}
\label{nonlin5}
F(\eta,\theta)&=&\frac{\gamma}{4}\epsilon^{4/3} K^4
\sinh^2\eta-\frac{\epsilon K^3}{a n \mu g}
\cos\bigg(\frac{\theta}{2}\bigg)
\cos\bigg(\frac{\eta}{2}\bigg)
\sinh^2\bigg(\frac{\eta}{2}\bigg)
\notag\\ && +\frac{\epsilon^{2/3} K^2}
{4n^2a^2 \mu^2\gamma g^2}\bigg[\sinh^2
\bigg(\frac{\eta}{2}\bigg)\cos^2\bigg(\frac{\theta}{2}\bigg)
+\frac{2}{3K^2}\sin^2\bigg(\frac{\theta}{2}\bigg)\bigg].
\end{eqnarray}
Here we have defined

\begin{equation}
\label{nonlin6}
U(\eta,\theta)=2\sigma(\eta,\theta)-\gamma k(\eta),
\end{equation}

\begin{equation}
\label{nonlin7}
g(\eta,\theta)=1+\frac{\epsilon}{\mu}
\cosh\bigg(\frac{\eta}{2}\bigg)\cos\bigg(\frac{\theta}{2}\bigg).
\end{equation}

Compactifying the warped deformed conifold throat via attaching
it to the compact Calabi-Yau space at some finite radius $r_{\text{UV}}$
requires K\"ahler modulus stabilization. This involves integrating 
out $\sigma$ by the assumption that it evolve adiabatically while 
remaining in its instantaneous minimum $\sigma_*(\eta,\theta)$ 
using the following equation:

\begin{equation}
\label{nonlin9}
\frac{\partial (V_F+V_{\text{uplift}})(\eta,\theta)}
{\partial\sigma}\bigg|_{\sigma_{\star}(\eta,\theta)}=0.
\end{equation}
Here we have added a D-term potential $V(\eta,\sigma)=D_{\text{uplift}}
/U(\eta,\sigma)^2$ which is needed to uplift KKLT AdS minimum to a dS minimum.
The uplifting can be sourced by distant anti-D3-branes or wrapped D7-branes. 
The instantaneous minimum of $\sigma$ is denoted by $\sigma_*$ including a 
shift due to its coordinate dependence induced by adding a mobile D3-brane 
to the compactification. The functional form of $\sigma_*$ can be determined
by the numerical solution of the transcendental equation (\ref{nonlin9}),
which we will solve in the next section. Before turning to numerical 
computation, we need to look at the minimum of the D3-D7 potential 
which specifies $\sigma_F$ and $D_{\text{uplift}}$ and constrains 
the rest of parameters on which these depend. 

The critical value $\sigma_F$ of the K\"ahler modulus before uplifting
is determined by the condition $D_{\sigma}W|_{\eta=0,\,\theta=0,\,
\sigma_{F}}=0$, or equivalently \cite{Kachru:2003aw},

\begin{equation}
\label{nonlin8}
e^{a\sigma_{F}}=\frac{|A_{0}|}{|W_{0}|}\bigg(1+\frac{2}{3}a\sigma_{F}\bigg)
g(0,0)^{1/n}\;\;\;\ \Rightarrow \;\;\;\ \frac{\partial V_F}{\partial\sigma}
\bigg|_{\sigma_F}=0,
\end{equation}
where $g$ is given by (\ref{nonlin7}). We may write this in the form:

\begin{eqnarray}
\label{1_g77}
\sigma_{F}&=&\frac{1}{a}\log\bigg[\frac{|A_{0}|}{|W_{0}|}
\bigg(1+\frac{2}{3}a\sigma_{F}\bigg)\bigg(1+\frac{\epsilon}{\mu}\bigg)^{1/n}\bigg].
\end{eqnarray}
From Eq.\,(\ref{nonlin4}) and Eq.\,(\ref{nonlin8}) we note that

\begin{equation}
V_F(0,\sigma_F)=-\frac{3a^2\kappa^2W_0^2}{2\sigma_F(3+2a\sigma_F)^2}.
\end{equation}
The uplifting parameter is given by

\begin{equation}
s=\frac{V_{\text{uplift}}(0,\sigma_F)}{|V_F(0,\sigma_F)|}
\,\,\,\,\,\,\ \text{with}\,\,\,\,\,\,\ V_{\text{uplift}}(0,\sigma_F)
=\frac{D_{\text{uplift}}}{4\sigma_F^2}.
\end{equation}
From this we obtain

\begin{equation}
\label{Dup}
D_{\text{uplift}}=\frac{6 s\, a^2\kappa^2W_0^2\sigma_F}{(3+2a\sigma_F)^2}.
\end{equation}
Here $1\le s \le 3$ to avoid decompactification, $a$ is determined by
the choice of $n$, $\kappa$ depends on the UV cut-off, and the value
of $\sigma_{F}$ can be derived from Eq.\,(\ref{1_g77}) once the set
of parameters \{$\epsilon$, $\mu$, $n$, $s$, $|A_{0}|$, $|W_{0}|$ \}
are specified. We also note that uplifting the KKLT AdS minimum to a
dS minimum introduces a small shift in the stabilized volume, 
$\sigma_0\equiv\sigma_F+\delta\sigma$. At the tip, we have \cite{Baumann:2007ah}:

\begin{equation}
\label{dsigF}
a\sigma_0\approx a\sigma_F+\frac{s}{a\sigma_F}.
\end{equation}
Here we note that both Eq.\,(\ref{1_g77}) and Eq.\,(\ref{dsigF}) which
give $\sigma_F$ and $\delta\sigma$, respectively, are derived from the
local minimum of the F-term  potential (\ref{nonlin4}) at the tip. The
critical value $\sigma_m$ of the K\"ahler modulus away from the tip and
its shift $\delta\sigma_m$ can be derived from the global minimum of 
the F-term potential  (\ref{nonlin4}). By computing the first and the 
second derivatives of the F-term potential $V_F$ with respect to $\sigma$,
it is straightforward to show that the global minimum of $V_F$ requires:

\begin{equation}
\label{sigma_m}
\sigma_m=\frac{1}{a}\Bigg[g(\eta,\theta)^{1/n}\frac{|A_{0}|}{|W_{0}|}
\bigg(1+\frac{aU}{3}+2aF\frac{aU+2}{aU+4}\bigg)\Bigg].
\end{equation}
Under the assuption that $V_D$ only shifts the minimum by a small amount gives:

\begin{eqnarray}
V'(\sigma_m+\delta\sigma)&=&V_D'(\sigma_m+\delta\sigma)
+V_F'(\sigma_m+\delta\sigma)\simeq V'_D(\sigma_m)+
\delta\sigma V''_F(\sigma_m)=0 \\ \Rightarrow\;\;\;\;\;\;\;\;\;\;\
 \label{deltasigmam}\delta\sigma &=&-\frac{V'_D(\sigma_m)}{V''_F(\sigma_m)}
=\frac{4V_D(\sigma_m)}{U_m V_F(\sigma_m)}\frac{V(\sigma_m)}{V''(\sigma_m)}.
\end{eqnarray}
We would like to remark here that in most of the brane inflation 
models in the literature (e.g. see \cite{Chen:2008ai,Baumann:2007ah})
the perturbative expansion of the potential is analysed in either 
the UV or the IR region. In these models $\epsilon$ and $\mu$ are 
considered to have similar order magnitude. However, we note that 
a consistent expansion along the entire throat including both the
UV and IR regions requires a large hierarchy between $\epsilon$
and $\mu$. In particular, the final piece of our potential term
$F(\eta,\theta)$ given by Eq.\,(\ref{nonlin5}) scales as 
$\epsilon^{-4/3}(\epsilon/\mu)^2$. Unless the hierarchy between 
$\epsilon$ and $\mu$ is large this term will dominate the potential
and destabilize the vacuum expansion. Moreover, the key point here
is that in our DBI brane inflation set up the full potential contains
a large leading order mass term for the radial coordinate and in order
to keep the adiabatic approximation (\ref{nonlin9}) valid the hierarchy
between $\epsilon$ and $\mu$ has to be large, so that the mass generated
for the K\"ahler modulus is much larger than the Hubble rate (see Section 4).

As mentioned above, apart from the shift induced by uplifting, $\delta\sigma$,
the addition of a mobile D3-brane in the compactification induces a further 
shift in the K\"ahler modulus that depends on the coordinates of the brane, 
$\sigma_*(\eta,\theta)$. Thus the nonperturbatively generated D3-D7 potential
about a dS minimum takes the form:

\begin{eqnarray}
\label{nonlin10}
V_{F}+V_{D}&=&\frac{2\kappa^2a_{n}^2|A_{0}|^2e^{-2a_{n}\sigma_*(\eta,\theta)}}
{U[\eta,\sigma_*(\eta,\theta)]^2}|g(\eta,\theta)|^{2/n}\notag\\ &&
 \times\bigg\{\frac{U[\eta,\sigma_*(\eta,\theta)]}{6} +\frac{1}{a_{n}}
\bigg(1-\frac{|W_{0}|}{|A_{0}|}\frac{e^{a_{n}\sigma_*(\eta,\theta)}}
{g(\eta,\theta)^{1/n}}\bigg)+F(\eta,\theta)\bigg\}\notag\\ && 
+\frac{D_{\text{uplift}}}{U[\eta,\sigma_*(\eta,\theta)]^2}.
\end{eqnarray}
Hence the total potential takes the form:

\begin{eqnarray}
T_3 V &=&  T_3 \left( \frac{1}{2} m_0^2 \; 
[r(\eta)^2 + c_2 K(\eta) \sinh \eta \cos\vartheta]+V_0 \right )
\notag\\ &&+V_F+V_D.
\label{D3pot}
\end{eqnarray}
Here $D_{\text{uplift}}$ is given by (\ref{Dup}), the constant
$V_0$ is chosen so that the global minimum of $V$ is $V=0$, 
and $c_2$ is an arbitrary constant. Because $c_2$ multiplies
a solution to a free Laplace equation, it is not fixed. 
For a self-consistent expansion, we expect $c_2$ to be smaller,
or of a magnitude comparable with other terms in the potential.

Finally, to derive the explicit form of the D3-brane equations
of motion on the deformed conifold for the simplest case including
only one angular direction for the potential (\ref{D3pot}) first 
note that a simple $S^3$ round metric on the deformed conifold can
be obtained from (\ref{dcmetric}) as:

\begin{eqnarray}
ds^2&=&A(\eta)d\eta^2+B(\eta)d\theta^2,
\end{eqnarray}
where

\begin{equation}
A(\eta)=\frac{\epsilon^{4/3}}{6 K(\eta)^2},\;\;\;\;\;\;\
B(\eta)=\frac{\epsilon^{4/3} K(\eta)}{4}\big[\cosh(\eta/2)
+\sinh(\eta/2)\big].
\end{equation}
The brane equations of motion can then be derived upon cross 
elimination from (\ref{EqfieldsNew}) in the following form:

\begin{eqnarray}
\label{nonlineq1}
\ddot{\eta}&=& -\frac{3H}{\gamma_{\text{DBI}}^2}\dot{\eta}+
\frac{h^{\prime}}{\gamma_{\text{DBI}} h}\dot{\eta}^2
(1-\gamma_{\text{DBI}})+\frac{h^{\prime}}{2h^2A}
(\gamma_{\text{DBI}}^{-1}-1)^2\notag\\ && -\frac{1}{2A}
(A^{\prime}\dot{\eta}^2-B^{\prime}\dot{\theta}^2)
+h\dot{\theta}\dot{\eta}\frac{V_{\theta}}{\gamma_{\text{DBI}} T_{3}}
-(1-hA\dot{\eta}^2)\frac{V_{\eta}}{\gamma_{\text{DBI}} A T_{3}}
\notag\\ && -\dot{\eta}\dot{\theta}(1-\gamma_{\text{DBI}}^{-1}
)\frac{h_{\theta}}{h},
\end{eqnarray}

\begin{eqnarray}
\label{nonlineq2}
\ddot{\theta} &=& -\frac{3H\dot{\theta}}{\gamma_{\text{DBI}}^{2}}
+(1-\gamma_{\text{DBI}})\dot{\theta}\dot{\eta}\frac{h^{\prime}}
{\gamma_{\text{DBI}} h}-\dot{\theta}\dot{\eta}\frac{B^{\prime}}
{B}+h\dot{\theta}\dot{\eta}\frac{V_{\eta}}{\gamma_{\text{DBI}} 
T_{3}}\notag\\ && -(1-hB\dot{\theta}^2)\frac{V_{\theta}}
{\gamma_{\text{DBI}} BT_{3}} -(1-\gamma_{\text{DBI}}^{-1})
\bigg[\dot{\theta}^2-\frac{(1-\gamma_{\text{DBI}}^{-1})}
{2hB}\bigg]\frac{h_{\theta}}{h}.
\end{eqnarray}
Here we note that in the absence of non-linear corrections including
the contribution of the D3-D7 potential and perturbations of the 
warp factor, the potential (\ref{D3pot}) and Eqs.\,(\ref{nonlineq1})
\,-\,(\ref{nonlineq2}) reduce to the D3-brane potential and equations
of motion including only linearized corrections studied in 
\cite{Gregory:2011cd}. We also note that in the slow-roll regime 
$\gamma_{\text{DBI}}\simeq 1$ and the perturbations of the warp factor
have no effect. However, since we are neither slow-rolling nor 
restricting ourselves to the linearized case, we need to take into 
account the perturbations of the warp factor which amount superpotential
corrections in the D3-D7 potential. For the warped deformed conifold, 
the perturbations of the warp factor in the tip region can be expressed
as \cite{Pufu:2010ie}

\begin{eqnarray}
\label{deltah1}
\delta h&=& -(2\pi)^4g_{s}p(\alpha^{\prime})^2 \mathcal{G}(\eta,y_4)
\notag\\ &=& \frac{16\pi\cdot 3^{2/3}g_{s}p(\alpha^{\prime})^2}
{(2\epsilon)^{8/3}\,\cdot\eta}\bigg[1+2\sqrt{2}y_4+6y_4^2 
+8\sqrt{2}y_4^3-\cdots \bigg]\;\;\ \text{with}\;\;\ \\ \label{y11}
 y_{4}&=&\bigg[\cos\bigg(\frac{\theta_{1}}{2}\bigg)
\sin\bigg(\frac{\theta_{2}}{2}\bigg)e^{\frac{i(\varphi_{1}-\varphi_{2}
+\psi)}{2}}-\sin\bigg(\frac{\theta_{1}}{2}\bigg)\cos\bigg(\frac{\theta_{2}}
{2}\bigg)e^{\frac{-i(\varphi_{1}-\phi_{2}-\psi)}{2}}\bigg].
\end{eqnarray}
Here $\mathcal{G}(\eta,y_4)$ is the Green's function (expanded in the 
eigenfunctions of the Laplacian) on the deformed conifold given by 
(\ref{1_g26}), $y_4$ comes from (\ref{y4}) and $p$ specifies the number 
of mobile D3-branes, which we take to be $p=1$. By the first angular 
condition in (\ref{nonlin2}) along an $S^3$ ($\theta_1=0$) for relations
(\ref{deltah1}) and (\ref{y11}) we have
\begin{eqnarray}
\label{deltahdc}
\delta h &=&\frac{16\pi\cdot 3^{2/3}g_{s}p(\alpha^{\prime})^2}
{(2\epsilon)^{8/3}\,\cdot\eta}\bigg[1+2\sqrt{2}y_4+6y_4^2 
+8\sqrt{2}y_4^3-\cdots \bigg]\;\;\ \text{with}\;\;\ \\
y_{4}&=&-\frac{1}{\sqrt{2}}\sin\bigg(\frac{\theta}{2}\bigg).
\end{eqnarray}
In the off-tip region, where $r$ is large,  we may use the very first
line in (\ref{deltah1}) together with relations (\ref{H2})\,-\,(\ref{norm})
for the quantum numbers $j_1=j_2=R/2=1/2$ and $m_1=m_2=1/2$, and obtain 
$\delta h$ in the form

\begin{eqnarray}
\label{nonlin152}
\delta h &\simeq&-\frac{10 g_s p {\alpha'}^2}{\pi^2}
\cos^4\frac{\theta}{2}\bigg(\frac{1}{r^4}\bigg)
+\cdots\;\;\ \text{with}\;\;\ \\ r^3 &=& \epsilon^2 e^{\eta}.
\end{eqnarray}

\section{Inflationary solutions}

\begin{figure}[!ht]
\begin{center}
\epsfig{file=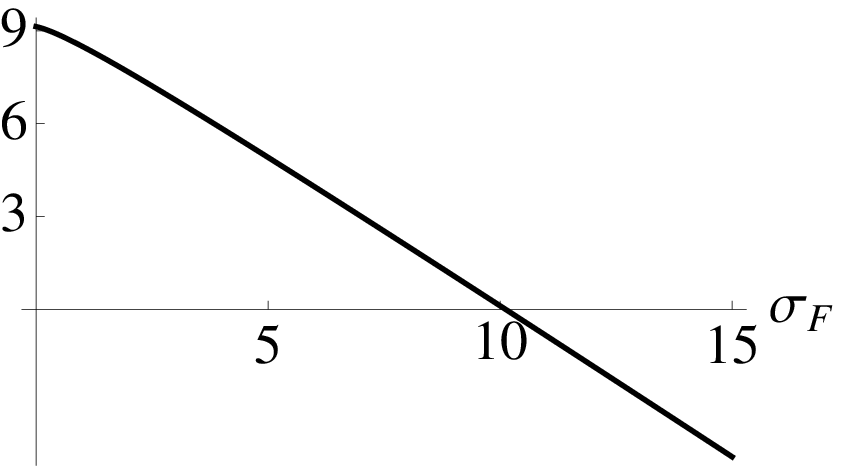,width=.45\textwidth}~~\nobreak
\epsfig{file=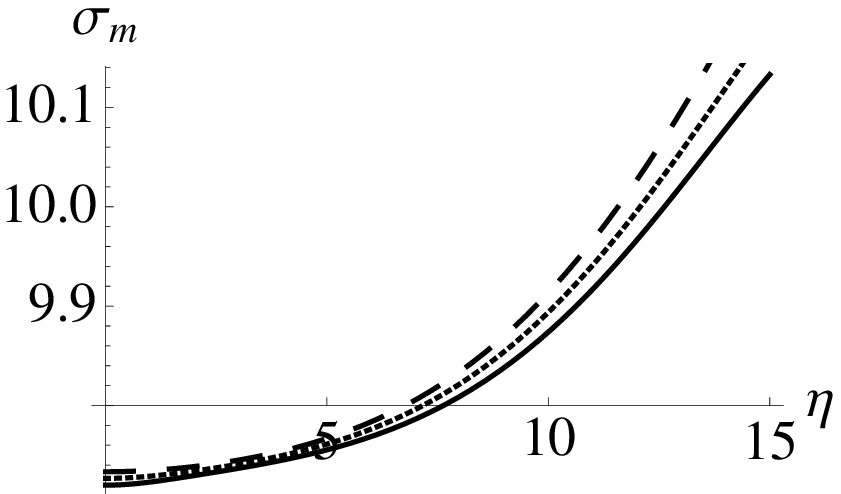,width=.45\textwidth}
\caption{The behaviour of $\sigma_F$ for choice of 
compactification parameters (4.1), and the behaviour
 of $\sigma_m$ for choice of parameters (4.1) with 
$\theta=0\,(\text{solid}), \pi/2\,(\text{tiny-dashed})
\,\, \text{and}\,\,\pi\,(\text{large-dashed})$.}
\label{fig:sigmafm}
\epsfig{file=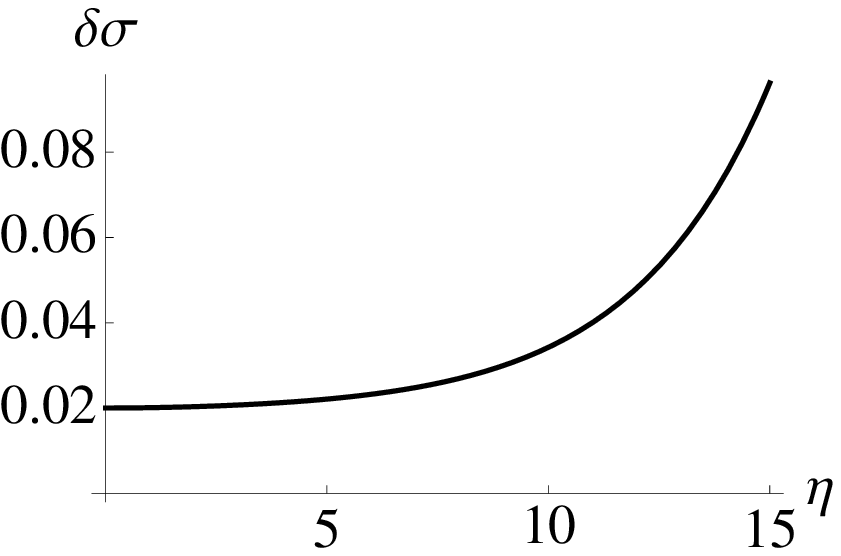,width=.45\textwidth}~~\nobreak
\epsfig{file=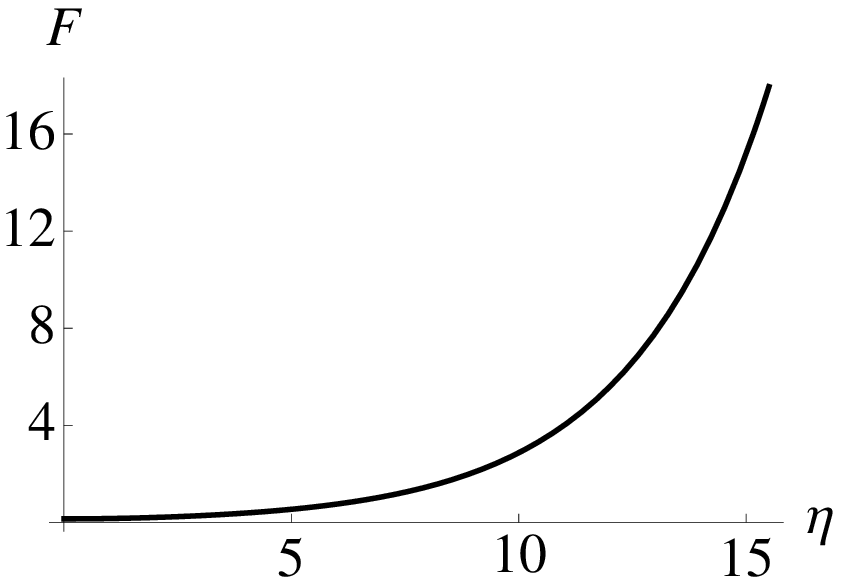,width=.45\textwidth}\\
\caption{The behaviour of the function $F$ and $
\delta\sigma$ for the choice of parameters (4.1) and
$\theta=\pi$.}
\label{fig:deltasigma2}
\epsfig{file=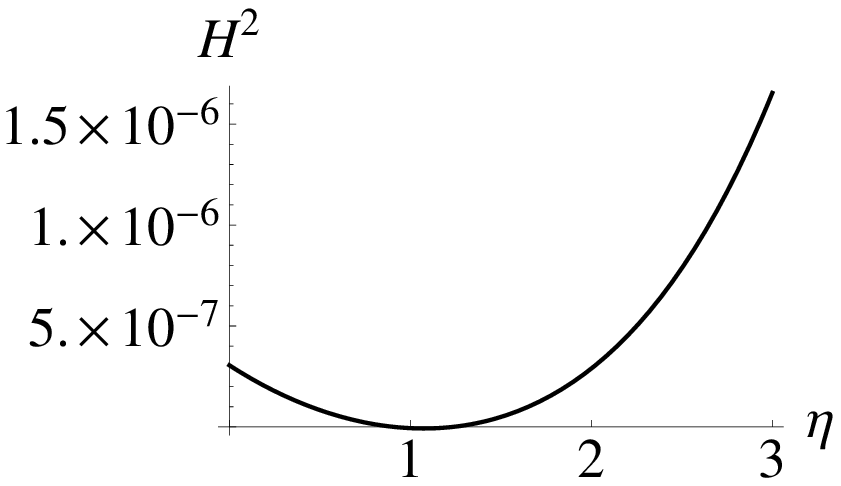,width=.45\textwidth}~~\nobreak
\epsfig{file=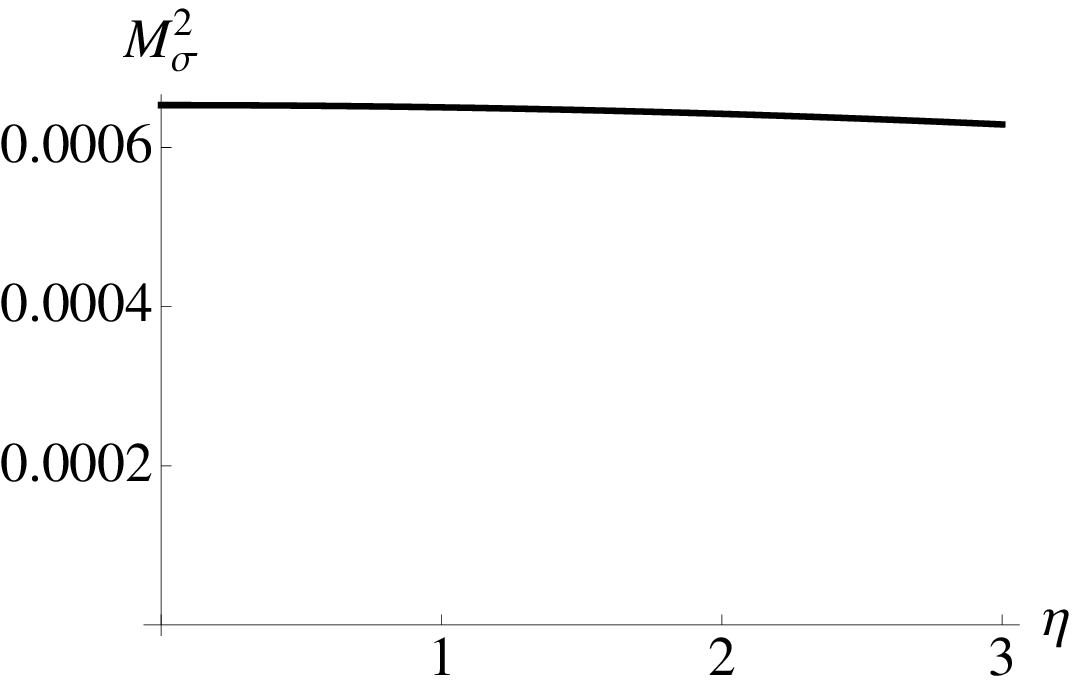,width=.45\textwidth}\\
\caption{The Hubble rate, $H^2\simeq V/3M_{\text{pl}}^2$
, and the $\sigma$ mass squared, $M_{\sigma}^2$ for the
 choice of parameters (4.1) and $\theta=\pi$.}
\label{fig:HS}
\end{center}
\end{figure}

In order to integrate the full brane equations of motion, we
 first need to specify a suitable choice of parameters and then
 compute the real part of the K\"ahler modulus that appears in th
e brane equations of motion. To choose a reasonable set of 
compactification parameters, we note the following points.
Firstly, we require a large hierarchy between $A_0$ and $W_0$ to 
guarantee large $\sigma_F$ which ensures suppressed 
$\alpha^{\prime}$-corrections. Secondly, we also need a large 
hierarchy between $\epsilon$ and $\mu$ in addition to the $A_0/W_0$ 
hierarchy to guarantee a valid perturbative expansion, $\delta\sigma\ll0$.
When both these hierarchies are turned on, $\sigma_{*}$ can be computed
within the adiabatic approximation from Eq.\,(\ref{nonlin9}).
We also remark from the literature that choosing a large value of
the UV-scale, $\eta_{\text{UV}}$, sets a large value for the Planck-mass
(e.g. see \cite{Gregory:2011cd}) in which case curvature corrections may
be omitted and non-linear corrections are dominated by IASD fluxes sourced
by moduli stabilizing wrapped D7-branes whose number is given by $n>1$. 
Furthermore, we remark that the supergravity solution requires
large $g_sM$ and the value of $s$ has to be chosen within the range
$1\le s\le 3$ to ensure a small positive cosmological constant 
and to avoid runaway decompactification.

In line with the above requirements we choose in our numerical analysis
the following specific set of compactification parameters:

\begin{align}
\label{tab:i}                                                                 
\eta_{\text{UV}}&=15, \;\;\ n=2,\;\;\ s=2,\;\;\ \epsilon=0.001\notag\\
g_sM &=100,\; \mu=5,\; W_0=29,\; A_0=25\times 10^{12}.
\end{align}
Inspection of Eq.\,(\ref{1_g77}) and Eq.\,(\ref{sigma_m}) shows that for
the choice of parameters (\ref{tab:i}) $\sigma_F$ and $\sigma_m$ scale
to large values (see Fig.\,\ref{fig:sigmafm}); inspection of 
Eq.\,(\ref{nonlin5}) and Eq.\,(\ref{deltasigmam}) shows that for the
choice of parameters (\ref{tab:i}) and $\theta=\pi$ the functions 
$\delta\sigma$ and $F$ scale to reasonably small values on the entire 
warped deformed conifold (see Fig.\,\ref{fig:deltasigma2}); for the choice
of parameters (\ref{tab:i}) the mass squared of the K\"ahler modulus, 
$M_{\sigma}^2\equiv V_{,\,\sigma\sigma}$, is much larger than the approximate
squared Hubble rate, $H^2\simeq V/3M_{\text{pl}}^2$ (see Fig.\,\ref{fig:HS}),
which guarantees an adiabatic expansion \cite{Baumann:2007ah,Panda:2007ie}. 
We note here that by decreasing these hierarchies much below their values 
(\ref{tab:i}) makes $M_{\sigma}^2$ and $H^2$ scale similarly which invalidates
the adiabatic approximation, in addition $\delta\sigma$ and $F$ become
large which invalidates the perturbative expansion.

\begin{figure}
\begin{center}
\epsfig{file=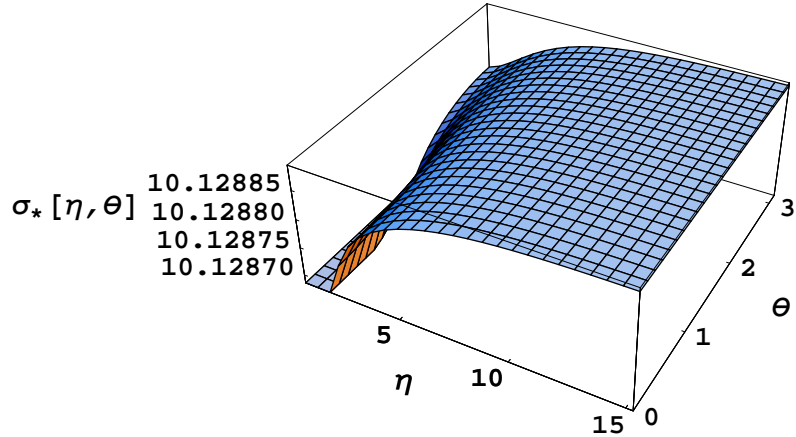,width=.7\textwidth}
\caption{The K\"ahler modulus, $\sigma_*(\eta,\theta)$, on the
entire supergravity background with the choice of compactification
parameters (4.1).}
\label{fig:KM}
\end{center}
\end{figure}

In the literature of brane inflation, the most common way of computing
the K\"ahler modulus, $\sigma$, is to adopt a semi\,-\,analytic approach
with the assumption that $\sigma$ evolves adiabatically (e.g. see
\cite{Baumann:2007np}). In this approach, $\sigma$ in $U(\eta, \sigma)$
is set to its large fixed value $\sigma_{0}$, and (\ref{nonlin9}) is 
treated as an equation in the variable $\exp[-a\sigma_{*}(\eta)]$. The
value of $\sigma_{*}$ is then obtained from Eq.\,(\ref{nonlin9}) by 
expanding in specific conical regions including either the IR or the UV
regions where the canonical inflaton (\ref{rks}) is given by its small
or large $\eta$ limit, respectively. This may be a good approximation 
but it gives only a qualitative understanding. 
Here we compute $\sigma_{*}$ by solving the transcendental equation 
(\ref{nonlin9}) numerically to obtain the exact value of $\sigma_*$ 
on the entire supergravity background. For the choice of parameters
(\ref{tab:i}) we show $\sigma_*$ in Fig.\,\ref{fig:KM}.

\begin{figure}
\begin{center}
\epsfig{file=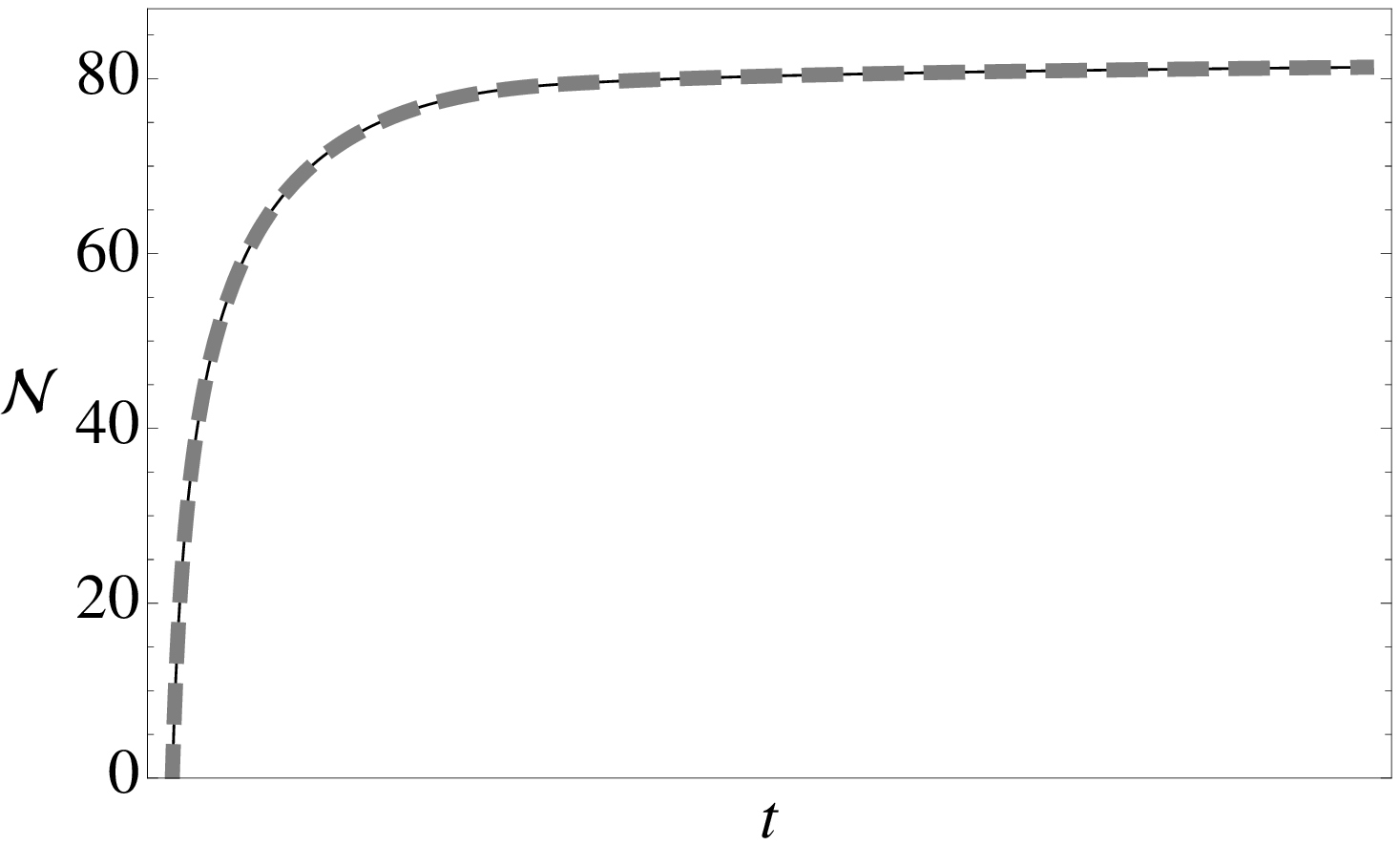,width=.7\textwidth}\\
\epsfig{file=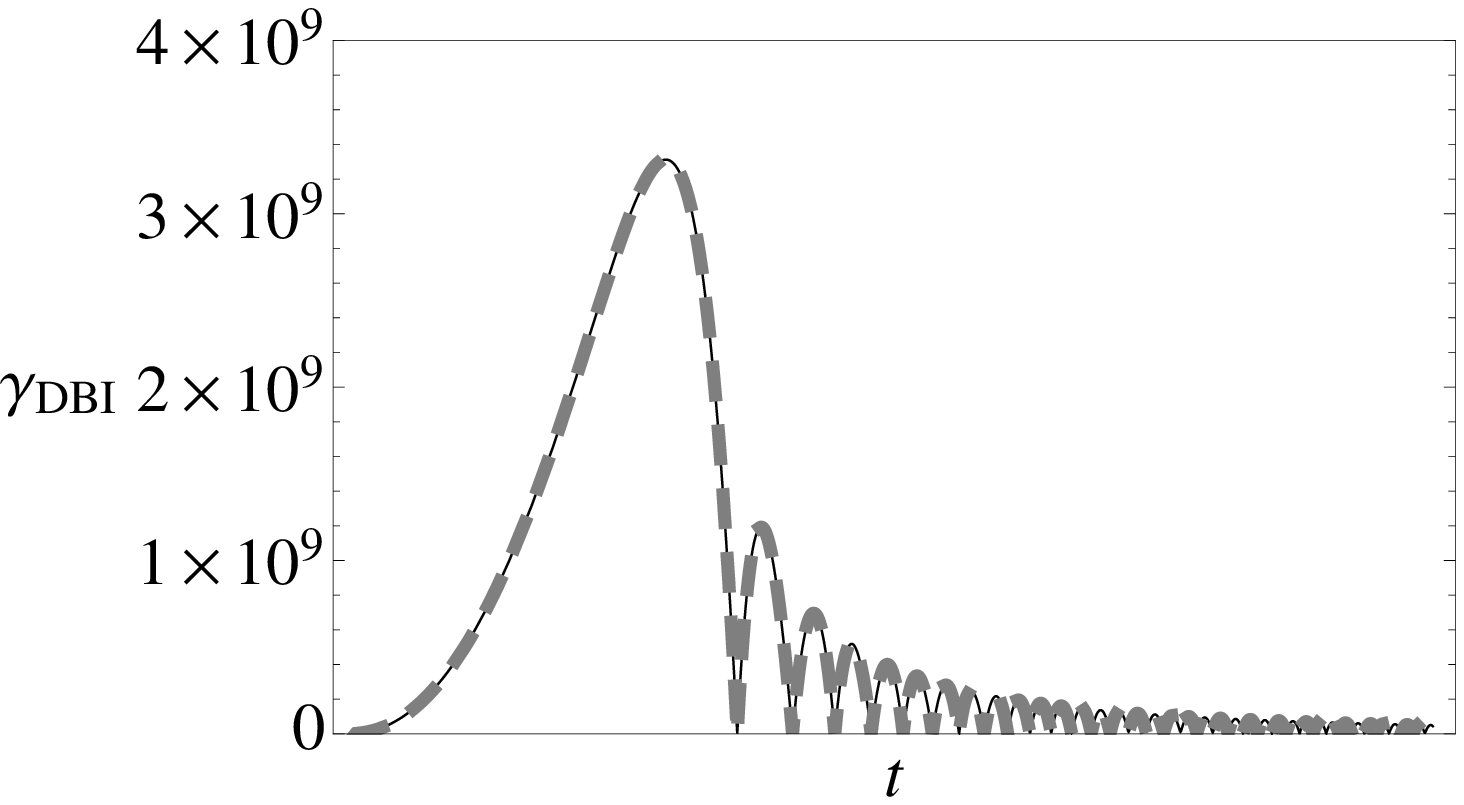,width=.7\textwidth}
\caption{The number of e-foldings, $\mathcal{N}$, and the
$\gamma$-factor, $\gamma_{\text{DBI}}$, with (gray-dashed) and
without (black-solid) non-linear harmonic dependent corrections
for the choice of compactification parameters (4.1).}
\label{fig:NeGammma}
\end{center}
\end{figure}

For the choice of parameters (\ref{tab:i}) and the numerically computed
K\"ahler modulus, we integrated the full D3-brane equations of motion,
Eqs.\,(\ref{nonhub1})\,-\,(\ref{EqfieldsNew}) with Eq.\,(\ref{EqfieldsNew})
given by Eqs.\,(\ref{nonlineq1})\,-\,(\ref{nonlineq2}), and our 
inflationary solution is displayed in Fig.\,\ref{fig:NeGammma}. 
The solution describes spiral brane motion at high speed in the warped
throat region of the compact Calabi-Yau space containing holomorphically
embedded wrapped D7-branes involved in (K\"ahler) moduli stabilization.
The conserved angular momentum is lifted by harmonic dependent corrections
from linearized as well as non-linear perturbations including contributions
from the D3-D7 potential and corrections of the warp factor. The brane 
accelerates along the radial and angular directions as it falls down the
throat from the UV end where it is attached to the compact Calabi-Yau 
space. Inflation ends when the brane reaches the IR location where the throat
smoothly closes off. For the choice of parameters and the numerically 
computed K\"ahler modulus, we integrated the brane equations of motion and
found that the inflationary solution is quite robust against harmonic dependent
corrections from the D3-D7 potential and corrections of the warp factor. In
particular, we found (as displayed in Fig.\,\ref{fig:NeGammma}) that harmonic
dependent corrections induced by the D3-D7 potential and perturbations of the
warp factor do not have the effect of increasing the number of e-foldings
and decreasing the $\gamma_{\text{DBI}}$-factor. This result differs from our
previous results \cite{Gregory:2011cd} in which harmonic dependent correction
to brane motion from linearized perturbations of the supergravity solution
increased the number of e-foldings compared to the number of e-foldings
produced by brane motion with conserved angular momentum (spinflation)
where no supergravity corrections and hence no harmonic dependence in 
brane motion is present.

We repeated the above computation for a various choices of initial 
conditions and compactification parameters and our findings are 
summarized as follows.
$\\$
$\bullet$ Decreasing $\epsilon$ below its considered value while 
keeping other parameters fixed tends to flatten the functional form
of $\sigma_*$ and changes its overall scale insignificantly. Also,
the decrease in $\epsilon$ leaves $\sigma_F$ unchanged.
$\\$
$\bullet$ Decreasing the value of $\mu$ while keeping other 
parameters fixed 
slightly decreases the value of $\sigma_F$ and changes the scale
of $\sigma_*$ by a minimal amount. Note that here we decrease $\mu$
by an amount, so that the hierarchy between $\epsilon$ and $\mu$ 
still remains large.
$\\$
$\bullet$ Increasing/decreasing $n$ while keeping other parameters 
fixed strongly impacts the $\sigma_F$ and the scale of $\sigma_*$. 
Also, increasing $n$ induces fluctuations in $\sigma_*$ at large $\eta$.
$\\$
$\bullet$ Increasing/decreasing the hierarchy between $A_0$ and $W_0$
while keeping other parameters fixed increases/decreases the value 
of $\sigma_F$ and slightly changes the scale of $\sigma_{*}$ but leaves
its overall shape unchanged. Here again we do not decrease the hierarchy
between $A_0$ and $W_0$ too much. Also, increasing the hierarchy between
$A_0$ and $W_0$ by taking $W_0$ much smaller than its considered value 
increases $\sigma_F$ and suppresses the uplifting contribution. This 
slightly changes the form of $\sigma_*$ near the origin.
$\\$
$\bullet$ Changing the set of compactification parameters and initial
conditions indicates that the number of e-foldings produced by spinflation
including non-linear harmonic dependent corrections depends more on the 
subset of compactification parameters $\{\epsilon, g_sM,\eta_{\text{UV}}\}$
than on the initial conditions and is insensitive to the choice of the 
remaning subset of compactification parameters $\{n,s,\mu,A_0,W_0\}$. 
Comparing with the number of e-foldings generated by spinflation 
including only linearized harmonic dependent corrections each time when
varying the set of parameters and initial conditions shows no differece
in the number of e-foldings (as Fig.\,\ref{fig:NeGammma}).
$\\$
$\bullet$ The $\gamma_{\text{DBI}}$-factor produced by spinflation 
including non-linear harmonic dependent corrections is insensitive
to the choice of compactification parameters and initial conditions.
Comparing with the $\gamma_{\text{DBI}}$-factor produced by spinflation
with only linearized harmonic dependent corrections each time when 
varying the set of parameters and initial conditions shows no differece
in the $\gamma_{\text{DBI}}$-factors (as Fig.\,\ref{fig:NeGammma}).

The above findings show that the inflationary solution is quite robust
against harmonic dependent corrections from the D3-D7 potential and
perturbations of the warp factor not just for the specific choice of
compactification parameters (\ref{tab:i}) but for a very large set of
consistent parameters.

\section{Summary and conclusions}

In this paper we studied brane inflation in a warped string 
compactification incorporating the effects of moduli stabilization
and backreaction from UV-deformations of the warped throat geometry.
The focus of our paper was on DBI brane inflation in the warped 
deformed conifold with a UV/IR consistent perturbative expansion
around the noncompact ISD solution. The perturbations were dominated
by IASD fluxes sourced by moduli stabilizing wrapped D7-branes.

We computed the D3-brane potential on the entire deformed conifold
including non-linear corrections from the flux induced potential 
in ten-dimensional supergravity which equals the nonperturbatively
generated D3-D7 brane potential in four-dimensional supergravity.
For a simple choice of a trajectory on the deformed conifold, we
integrated out the K\"ahler modulus and reduced the D3-brane potential
to a simple two-field potential depending on radial and harmonic
directions of the deformed conifold. We integrated out the K\"ahler
modulus by full numerical computation determining its exact functional
form on the entire supergravity background including both the IR and 
UV regions. We found that a UV/IR consistent perturbative expansion in
the supergravity potential with the K\"ahler modulus integrated out 
within the adiabatic approach in DBI inflation requires certain 
hierarchies of scales that determine the set of compactification 
parameters different from those in slow-roll models. 

For the consistent choice of parameters and the numerically computed
K\"ahler modulus, we integrated the D3-brane equations of motion in
the warped deformed conifold with harmonic dependence from the D3-brane
potential and perturbations of the warp factor. We found that our 
numerical solutions are quite robust against non-linear perturbations
including harmonic dependent corrections from perturbations of the 
warp factor and the D3-D7 brane potential.
In particular, we found that harmonic dependent corrections from the
D3-D7 potential and perturbations of the warp factor do not have the
effect of increasing the number of e-foldings and decreasing the 
$\gamma_{\text{DBI}}$-factor. We therefore conclude that the most leading
order harmonic dependent correction to brane (spin)inflation comes from
the linearized corrections analysed in \cite{Gregory:2011cd} with the
level of non-Gaussianity remaining large.

Our analysis can be extended in several ways. One direction would be to
consider different embedding functions for D7-branes with different 
trajectories on the deformed conifold and see how this may affect the 
inflationary solutions. Despite the fact that different embedding functions
for D7-branes and different trajectories on the deformed conifold modify 
the functional form of the supergravity potential, the K\"ahler modulus 
and perturbations of the warp factor, we expect this to have a subdominant
effect on the number of e-foldings and the $\gamma_{\text{DBI}}$-factor.
It would also be interesting to consider the possibility of a dynamical
K\"ahler modulus (instead of stabilized) \cite{Panda:2007ie}, and
integrate the brane equations of motion for a less restricted choice 
of parameters. Since taking the K\"ahler modulus field dynamical adds
to the number of brane equations of motion, we expect this to have a
less trivial impact on the inflationary solutions, though we do not 
expect this to decrease the $\gamma_{\text{DBI}}$-factor by an 
appreciable amount. 

The other, perhaps more interesting way of extending our analysis is
to consider further corrections to the inflaton action and analyse 
in detail the effects of cosmological perturbation theory. One particular
correction comes from the contribution of the flux induced potential of
harmonic type \cite{Baumann:2010ll}.
The flux induced D3-D7 potential that we considered in our inflationary
analysis came from the holomorphic solution of the noncompact 
supergravity equation of motion which described non-linear 
perturbations around the ISD solution. Since the general 
solution should be harmonic rather than just holomorphic it would
be necessary to include such harmonic contributions which have not
been computed for the deformed conifold to date. 

Another correction to the inflaton action arises from departures
of the noncompact limit \cite{Baumann:2010ll} considered in this
paper. Despite the fact that our Planck-mass was set large by the
UV-scale we considered, it would be interesting to consider possible
departures from the noncompact limit and compute further contributions
to the D3-brane potential from coupling to curvature corrections which
also induce harmonic dependence in brane motion \cite{Baumann:2010ll}.
In particular, coupling to the Ricci-scalar introduces a non-minimal 
coupling in the DBI action which corrects the $\gamma_{\text{DBI}}$-factor
and may have the capacity of decreasing the level of non-Gaussianity 
significantly \cite{vandeBruck:2010yw}. It would be interesting to confirm
this result in the concrete supergravity set up considered in this paper
and investigate its implications for cosmological perturbations in more
detail \cite{Weller:2011ey} in our framework. 

Finally, it would be important to confirm whether in our supergravity
set up multifield effects (e.g. from phase transition) induced by an
instability along harmonic directions \cite{Kidani:2012jp} do have the
capacity to evade stringent constraints in cosmological perturbations
for single field inflationary models. Moreover, it would be very
interesting, though formidable, to consider our DBI brane inflation
model along a trajectory on the deformed conifold depending on all six
directions and extract the full multifield effects for cosmological 
perturbations in a UV/IR consistent expansion and make contact with 
some of the results obtained in \cite{McAllister:2012am} by taking the
singular conifold limit.  We shall leave the investigation of these for
the future.

\section*{Acknowledgment}

I am very grateful to Prof. Ruth Gregory for useful discussions on
this paper and for collaboration on previous work.

\end{document}